\documentclass{article}
\usepackage{epsfig}

\newcommand{\lsim}{\mbox{$\stackrel{\scriptstyle <}
{\scriptstyle \sim}$}}
\newcommand{\gsim}{\mbox{$\stackrel{\scriptstyle >}
{\scriptstyle \sim}$}}

\newcommand\decay{%
\;\rule[0.5ex]{0.4pt}{1ex}\hspace*{-4pt}\rightarrow}

\begin{document}

\title{News on Neutrino Oscillations and Neutrino Masses
\thanks{Invited talk
    at the XXVIII International Symposium on Multiparticle Dynamics,
    Delphi, Greece, September 1998;
    MPI Preprint {\bf MPI-PhE/98-15}}}

\author{Norbert Schmitz\\
Max-Planck-Institut f\"ur Physik, F\"ohringer Ring 6\\
80805 M\"unchen, Germany\\ E-mail: nschmitz@mppmu.mpg.de} 


\maketitle
\begin{abstract}
Recent results on neutrino
  oscillations and neutrino masses are presented. 
The most exciting news are the Super-Kamiokande measurements of
  atmospheric neutrinos, which show evidence for the neutrinos being
  massive. Various possible schemes for the neutrino masses are discussed.
\end{abstract}

\section{Introduction}
This report has been given at the request of the organizers of the
Delphi symposium. It summarizes the most exciting news on neutrino
oscillations and neutrino masses, 
as they were presented at the Neutrino '98 Conference 
($\nu$'98) at Takayama (Japan) in June 1998, at the Ringberg Euroconference
(New Trends in Neutrino Physics) in May 1998, and in various recent
publications. 

One of the fundamental questions in particle physics is as to whether
neutrinos have a mass ($m_\nu > 0$, massive neutrinos) or are exactly
massless (like the photon). For the following reason this question is
directly related to the more general problem whether there is new
physics beyond the Standard Model (SM): In the minimal SM 
neutrinos have a fixed helicity, always $H (\nu ) = -1$ and $H (\bar{\nu}) = +1$. This implies $m_\nu = 0$, since only massless particles can be eigenstates of the helicity operator. $m_\nu > 0$ would therefore transcend the simple SM.

Direct kinematic measurements of neutrino masses have so far yielded only rather loose upper limits, the present best values being \cite{r1}:
\begin{equation}
\label{e1}
\begin{array}{l@{\ <\ }l@{\hspace{0.5cm}}l}
m (\nu_e ) & 15\ {\rm eV}& \mbox{(from tritium $\beta$-decay)}\\
m (\nu_\mu ) & 170\ {\rm keV}\ \ \ \mbox{(90\% CL)} &
                            \mbox{(from $\pi^+$ decay)}\\
m (\nu_\tau )& 18.2\ {\rm MeV}\ \mbox{(95\% CL)} &
                             \mbox{(from $\tau$ decays).}
\end{array}
\end{equation}
These limits will very likely not be improved considerably in the future.

Access to much smaller mass values is provided by neutrino oscillations
\cite{r2}. They allow, however, to measure only differences of masses
squared, $\delta m^2_{ij} \equiv m^2_i - m^2_j$, rather than masses
directly. For completeness we summarize briefly the most relevant formulae
for neutrino oscillations in the vacuum, considering for simplicity only
two neutrino flavours $(\nu_a , \nu_b)$, e.g.\ $(\nu_e , \nu_\mu )$
(two-flavour formalism). The generalisation to three or more flavours is
straight-forward in principle, but somewhat more involved in 
practice \cite{r2}.

\begin{figure}
\begin{center}
\epsfig{file=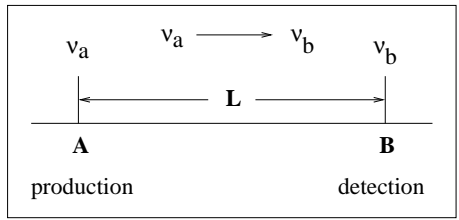,width=6cm}
\end{center}
\caption{Scheme of a neutrino oscillation experiment.}
\label{fig1}
\end{figure}

The two flavour eigenstates $(\nu_a , \nu_b )$ are in general related to
the two mass eigenstates $(\nu_1 , \nu_2 )$ with masses $(m_1 , m_2 )$
by a unitary transformation:
\begin{equation}
\label{e2}
\left (
\begin{array}{c}
\nu_a\\
\nu_b
\end{array}
\right )
=
\left (
\begin{array}{c@{\hspace{0.3cm}}c}
\cos \theta & \sin \theta\\
-\sin \theta & \cos \theta
\end{array}
\right ) \cdot \left (
\begin{array}{c}
\nu_1\\
\nu_2
\end{array}
\right )
\end{equation}
where $\theta$ is the mixing angle. If $m_1 \neq m_2$, the two eigenstates 
$(\nu_1 , \nu_2 )$ evolve differently in time, so that for $\theta \neq 0$ a given original linear superposition of $\nu_1$ and $\nu_2$ changes with time into a different superposition. This means that flavour transitions (oscillations) $\nu_a \to \nu_b$ and $\nu_b \to \nu_a$ can occur with certain time-dependent oscillatory probabilities. In other words 
(Fig.~\ref{fig1}): 
If a neutrino is produced (or detected) at A as a flavour eigenstate
$\nu_a$ (e.g.\ $\nu_\mu$ from $\pi^+ \to \mu^+ + \nu_\mu$), it is detected,
after travelling a distance (baseline) $L$, at B with a probability $P
(\nu_a \to \nu_b )$ as flavour eigenstate $\nu_b$ (e.g.\ $\nu_e$ in $\nu_e
n \to pe^-$). The probability $P (\nu_a \to \nu_b) = P (\overline{\nu}_a
\to \overline{\nu}_b) = P (\nu_b \to \nu_a)$ is given by
\begin{equation}
\label{e3}
\left.\begin{array}{ll}
P (\nu_a \to \nu_b )&
= \sin^2 2\theta \cdot \sin^2 \left (
{\displaystyle\frac{\delta m^2}{4}} \cdot
{\displaystyle\frac{L}{E}}\right )\\[2.5ex]
&= \sin^2 2\theta \cdot \sin^2 \left ( 1.267 
{\displaystyle\frac{\delta m^2}
{{\rm eV}^2} \cdot \frac{L/{\rm m}}{E/{\rm MeV}}}\right )
\end{array}\right\rbrace
\begin{array}{l}
{\rm for}\ \nu_a \neq \nu_b\\
({\rm flavour~change}\\
\nu_a \to \nu_b)
\end{array}
\end{equation}
\begin{displaymath}
\hspace*{-3.3cm}
P (\nu_a \to \nu_a )
= 1 - P (\nu_a \to \nu_b )\ \ \ \ \ \mbox{(survival of $\nu_a$)}
\end{displaymath}
where $\delta m^2 = m^2_2 - m^2_1$ and $E =$ neutrino energy.
Thus the probability oscillates when varying $L/E$, with $\theta$
determining the amplitude $(\sin^2 2\theta )$ and $\delta m^2$ the
frequency of the oscillation. The smaller $\delta m^2$, the larger $L/E$
values are needed to see oscillations, i.e.\ significant
deviations of $P (\nu_a \to \nu_b )$ 
from zero and of $P (\nu_a \to \nu_a )$ from unity. 
Notice the two necessary
conditions for $\nu$ oscillations: (a) $m_1 \neq m_2$ implying that not all neutrinos are massless, and (b) non-conservation of the lepton-flavour numbers. 

The masses $m (\nu_a )$ and $m (\nu_b )$ of the flavour eigenstates are
 expectation values of the mass operator, i.e.\ linear combinations of $m_1$ and $m_2$:
\begin{equation}
\label{e4}
\begin{array}{l}
m (\nu_a ) = \cos^2 \theta \cdot m_1 + \sin^2 \theta \cdot m_2\\[1ex]
m (\nu_b )= \sin^2 \theta \cdot m_1 + \cos^2 \theta \cdot m_2
\end{array}
\end{equation}
For a small mixing angle: $m (\nu_a) \approx m_1$, 
$m (\nu_b) \approx m_2$. For maximum mixing ($\theta = 45^\circ$):
$m (\nu_a) = m (\nu_b) = (m_1 + m_2)/2$.

\section{Atmospheric neutrinos}
The most exciting new results presented at $\nu$'98 come from
Super-Kamiokande on atmospheric neutrinos; they show convincing
evidence for neutrino oscillations and thus for massive neutrinos.

Atmospheric neutrinos are created when a
high-energy cosmic-ray proton (or nucleus) from outer space collides with a
nucleus in the earth's atmosphere, leading to an extensive air shower (EAS) by 
cascades of secondary
interactions. Such a shower contains many $\pi^\pm$ (and $K^\pm$) mesons (part
of) which decay according to
\begin{equation}
\label{e5}
\begin{array}{ll@{\hspace{1cm}}ll}
\pi^+ , K^+ \rightarrow & \mu^+ \nu_\mu &
\pi^- , K^- \rightarrow & \mu^- \overline{\nu}_\mu \\
&\decay e^+ \nu_e \overline{\nu}_\mu && \decay e^- \overline{\nu}_e
\nu_\mu \, ,
\end{array}
\end{equation}
yielding atmospheric neutrinos.
From (\ref{e5}) one would expect in an underground neutrino detector 
a number ratio of
\begin{equation}
\label{e6}
\frac{\mu}{e} \equiv \frac{\nu_\mu + \overline{\nu}_\mu}
{\nu_e + \overline{\nu}_e} = 2\, ,
\end{equation}
if all $\mu^\pm$ decayed before reaching the detector.
 This is the case only at rather low shower
energies whereas with increasing energy more and more $\mu^\pm$ survive
due to relativistic time dilation  (atmospheric
$\mu$). Consequently 
the expected $\mu /e$ ratio rises above 2 (fewer and fewer 
$\nu_e , \overline{\nu}_e$) with increasing $\nu$ energy. For quantitative
predictions Monte Carlo (MC) simulations,
which include also other (small) $\nu$ sources, have been performed
modelling the
air showers in detail and yielding the fluxes of the various neutrino
species ($\nu_e , \overline{\nu}_e , \nu_\mu , \overline{\nu}_\mu$) as a
function of the $\nu$ energy \cite{r42}. 
The various calculations agree on the
absolute $\nu$ fluxes only within $\sim 30\%$ whereas the agreement on the
flux ratio $\mu /e$ is much better, namely within  $\sim 5\%$.

Atmospheric neutrinos reaching the underground 
Super-Kamiokande
detector can be registered by neutrino reactions inside the detector, the
simplest and most frequent reactions being CC quasi-elastic scatterings:
\begin{equation}
\label{e7}
({\rm a})\mbox{\hspace{0.3cm}}
\begin{array}{l}
\nu_e n \to p e^-\\
\overline{\nu}_e p \to ne^+
\end{array}
\mbox{\hspace{0.8cm} (b)\hspace{0.3cm}}
\begin{array}{l}
\nu_\mu n \to p \mu^-\\
\overline{\nu}_\mu p \to n\mu^+ \, .
\end{array}
\end{equation}

\begin{figure}
\begin{center}
\epsfig{file=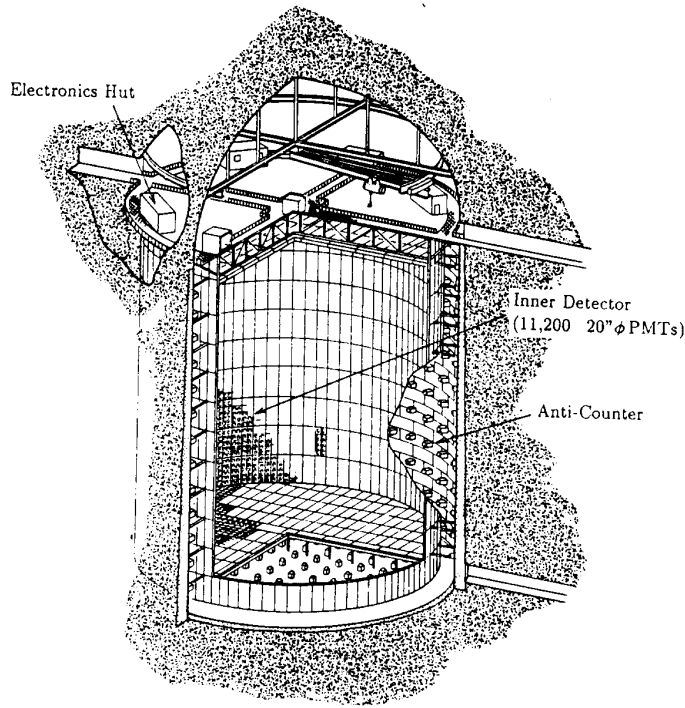,width=9.5cm}
\end{center}
\caption{Schematic view of Super-Kamiokande \protect\cite{r3}.}
\label{fig2}
\end{figure}

Super-Kamiokande
(Fig.~\ref{fig2}) \cite{r3}
is a big water-Cherenkov detector in the Ka\-mi\-o\-ka Mine (Japan) at a depth of
$\sim$ 1000 m. It consists of 50 ktons (50\,000 m$^3$) of 
ultrapurified water
in a cylindrical tank (diameter = 39 m, height = 41 m). The inner detector
volume of 32 ktons is watched by 11\,146 photomultiplier tubes (PMTs,
diameter = 20$''$) mounted on the volume's surface and providing a 40\%
surface coverage. The outer detector, which
vetos entering particles and tags exiting particles,
 is a 2.5 m thick water layer
surrounding the inner volume and looked at by 1885 smaller PMTs (diameter =
8$''$). 
A high-velocity charged particle passing through the water produces a
cone of Cherenkov light which is registered by the PMTs. The Cherenkov
image of a particle starting and ending inside the inner detector is a
ring, the image of a particle starting inside
 and leaving the inner detector is
a disk. A distinction between an $e$-like event (\ref{e7}a) and a
$\mu$-like event (\ref{e7}b) is possible 
(with an efficiency of $\gsim\ 98\%$)
from the appearance of the image:
an $e^\pm$ has an image with a diffuse, fuzzy boundary 
whereas the boundary of a $\mu^\pm$ image is sharp. The
observed numbers of $\mu$-like and $e$-like events give directly the
observed
ratio $(\mu /e)_{\rm obs}$ which is to be compared with the 
MC-predicted ratio $(\mu /e)_{\rm MC}$ (for no $\nu$ oscillations)
by computing the double ratio
\begin{equation}
\label{e8}
R = \frac{(\mu /e)_{\rm obs}}{(\mu /e)_{\rm MC}}\, .
\end{equation}
Agreement between observation and expectation implies $R = 1$.

The events are separated into fully contained events (FC, no track leaving
the inner volume) and partially contained events (PC, one or more tracks
leaving the inner volume). For FC events the visible energy $E_{\rm vis}$,
which is obtained from the pulse heights in the PMTs, is close to the $\nu$
energy. With this in mind, the FC sample is subdivided into sub-GeV events
($E_{\rm vis} < 1.33$ GeV) and multi-GeV events ($E_{\rm vis} > 1.33$ GeV).
In the multi-GeV range
 the $\nu$ direction is
determined as the direction of the Cherenkov-light cone, since at higher
energies the directions of the incoming $\nu$ and the outgoing charged
lepton are close to each other.

The double ratio $R$ has been measured previously by Kamiokande, the
smaller predecessor of Super-Kamiokande
 with 3.0 ktons of water, yielding \cite{r5}
\begin{equation}
\label{e9}
\begin{array}{r@{\hspace{0.3cm}}l}
R = 0.60 \pm {0.06 \atop 0.05}\ {\rm (stat.)}\ \pm 0.05\ {\rm (syst.)}&
\mbox{from 482 sub-GeV events}\\[1ex]
 = 0.57 \pm {0.08 \atop 0.07}\ {\rm (stat.)}\ \pm 0.07\ {\rm (syst.)}&
\mbox{from 233 multi-GeV events.}
\end{array}
\end{equation}
The new Super-Kamiokande
 result, based on larger statistics, is \cite{r6,r7}
\begin{eqnarray}
\label{e10}
R &=& 0.63 \pm 0.03 \ {\rm (stat.)}\ \pm 0.05\ {\rm (syst.)}
\ \ \ \mbox{from 2389 FC sub-GeV events}\nonumber \\[-1ex]
&&\\[-0.5ex]
&=& 0.65 \pm 0.05 \ {\rm (stat.)}\ \pm 0.08\ {\rm (syst.)}\ \ \ 
\mbox{from 520 FC multi-GeV events}\nonumber \\[-0.5ex]
&&\hspace{5.4cm}\mbox{and 301 PC events.}\nonumber
\end{eqnarray}
In all cases $R$ is significantly smaller than unity 
(atmospheric $\nu$ anomaly)
which is due, as it
turns out (see below), to a deficit of $\nu_\mu , \overline{\nu}_\mu$ and
not to an excess of $\nu_e , \overline{\nu}_e$
in $(\mu /e)_{\rm obs}$. A natural explanation of this deficit is that
some $\nu_\mu , \overline{\nu}_\mu$ have oscillated into 
$(\nu_e , \overline{\nu}_e)$ or $(\nu_\tau , \overline{\nu}_\tau)$
according to (\ref{e3}) before reaching the detector. However, 
the solution $\nu_\mu \to \nu_e$ of Kamiokande (not of Super-Kamiokande)
is ruled out by the CHOOZ experiment (see below) so that 
probably only  $\nu_\mu \to \nu_\tau$ remains.

This explanation has been evidenced by a study of the $\nu$ fluxes and of
$R$ as a function of the zenith angle $\Theta$ between the vertical
(zenith) and the $\nu$ direction. A $\nu$ with 
$\Theta \approx 0^\circ$ comes from above (down-going $\nu$) after
 travelling a distance of $L\ \lsim\ 20$ km (effective thickness of the
 atmosphere); a $\nu$ with $\Theta \approx 180^\circ$ reaches the detector
 from below (up-going $\nu$) after traversing the whole earth with $L
 \approx 13000$ km. 

\begin{figure}
\begin{minipage}[b]{4.8cm}
\epsfig{file=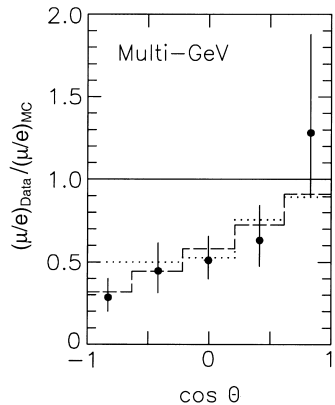,width=4.5cm}
\end{minipage}
\hfill
\begin{minipage}[b]{7cm}
\caption{Dependence of the ratio $R$, eq.~(\ref{e8}), 
on the zenith angle $\Theta$
  for multi-GeV events from Kamiokande. The dotted histogram shows the MC
  prediction including $\nu_\mu \leftrightarrow \nu_\tau$ oscillations with
  the parameters (\ref{e11}). The dashed histogram is for $\nu_\mu
  \leftrightarrow \nu_e$ oscillations \protect\cite{r5}.}
\label{fig3}
\end{minipage}
\end{figure}

Fig.~\ref{fig3} 
shows for multi-GeV events (for which the $\nu$ direction can be
determined) the dependence of $R$ on $\cos \Theta$ as measured by
Kamiokande \cite{r5}. Up-$\nu_\mu$ are seen to be missing ($R < 1$) whereas
down-$\nu_\mu$ appear in the expected frequency $(R \approx 1$). 
In terms of
$\nu$ oscillations this means that part of the up-$\nu_\mu$
have changed their flavour on their long way through the earth whereas for
down-$\nu_\mu$ $L$ is so small that $P (\nu_\mu \to \nu_e ,\nu_\tau)$,
eq.~(\ref{e3}), is practically zero. The $\nu_\mu$ deficit in 
Fig.~\ref{fig3} 
is the larger, the larger $\Theta$ and thereby $L$ is. A quantitative
oscillation analysis yielded as the best-fit parameters \cite{r5}
\begin{equation}
\label{e11}
\delta m^2 = 1.6 \cdot 10^{-2}\ {\rm eV}^2\, ,\ \ \sin^2 2\theta = 1.0
\ \ \mbox{for $\nu_\mu \leftrightarrow \nu_\tau$}\, .
\end{equation}
The fit (dotted histogram in  
Fig.~\ref{fig3})
describes the data points rather well.

\begin{figure}
\begin{minipage}[b]{4.5cm}
\epsfig{file=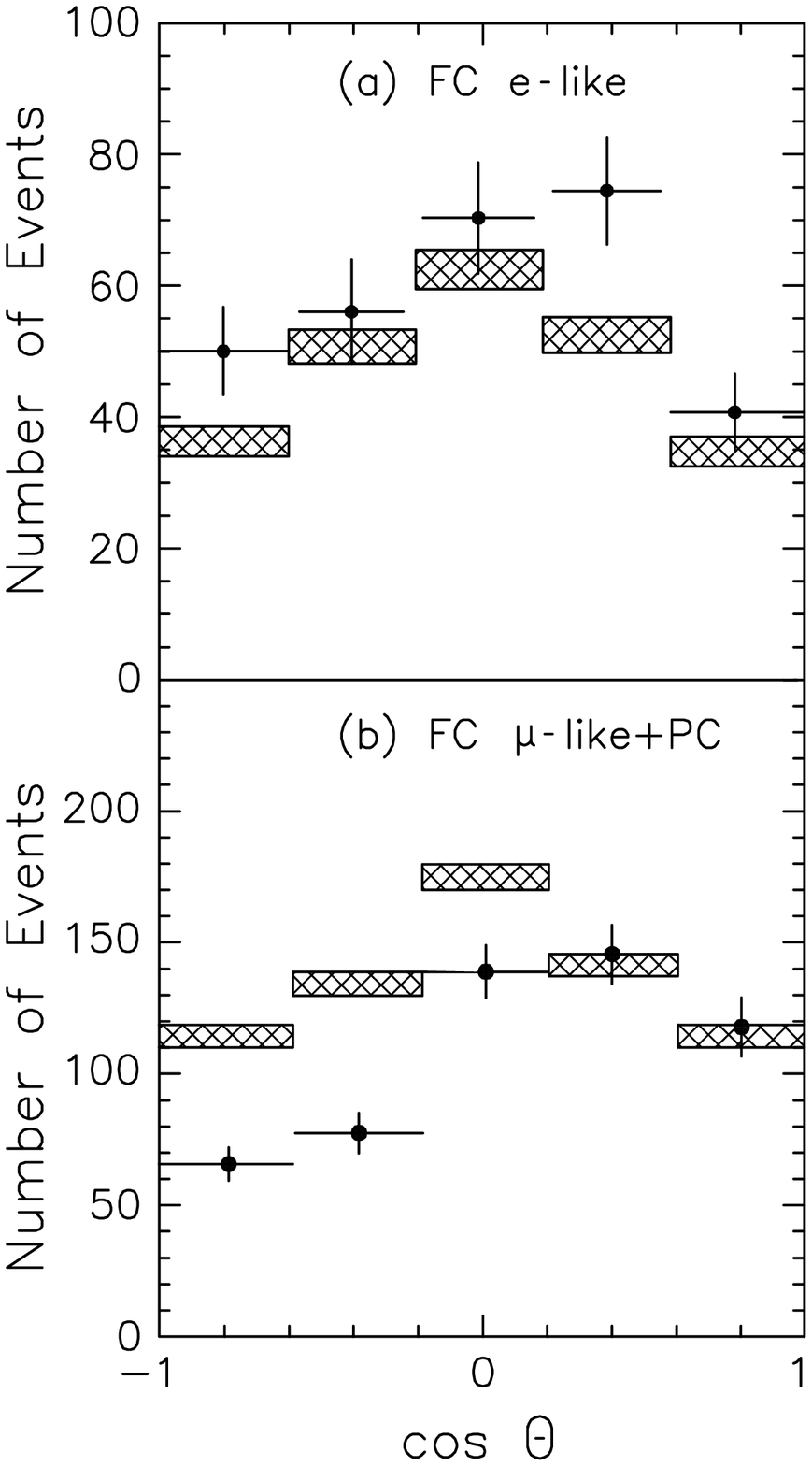,width=4.5cm}
\caption{Zenith-angle distribution of (a) FC $e$-like and (b) FC $\mu$-like
  + PC events in the multi-GeV range
from Super-Kamiokande. The points show the data, the
  rectangles show the MC prediction for no oscillations \protect\cite{r6}.}
\label{fig4}
\end{minipage}
\hfill
\begin{minipage}[b]{6cm}
\begin{center}
\epsfig{file=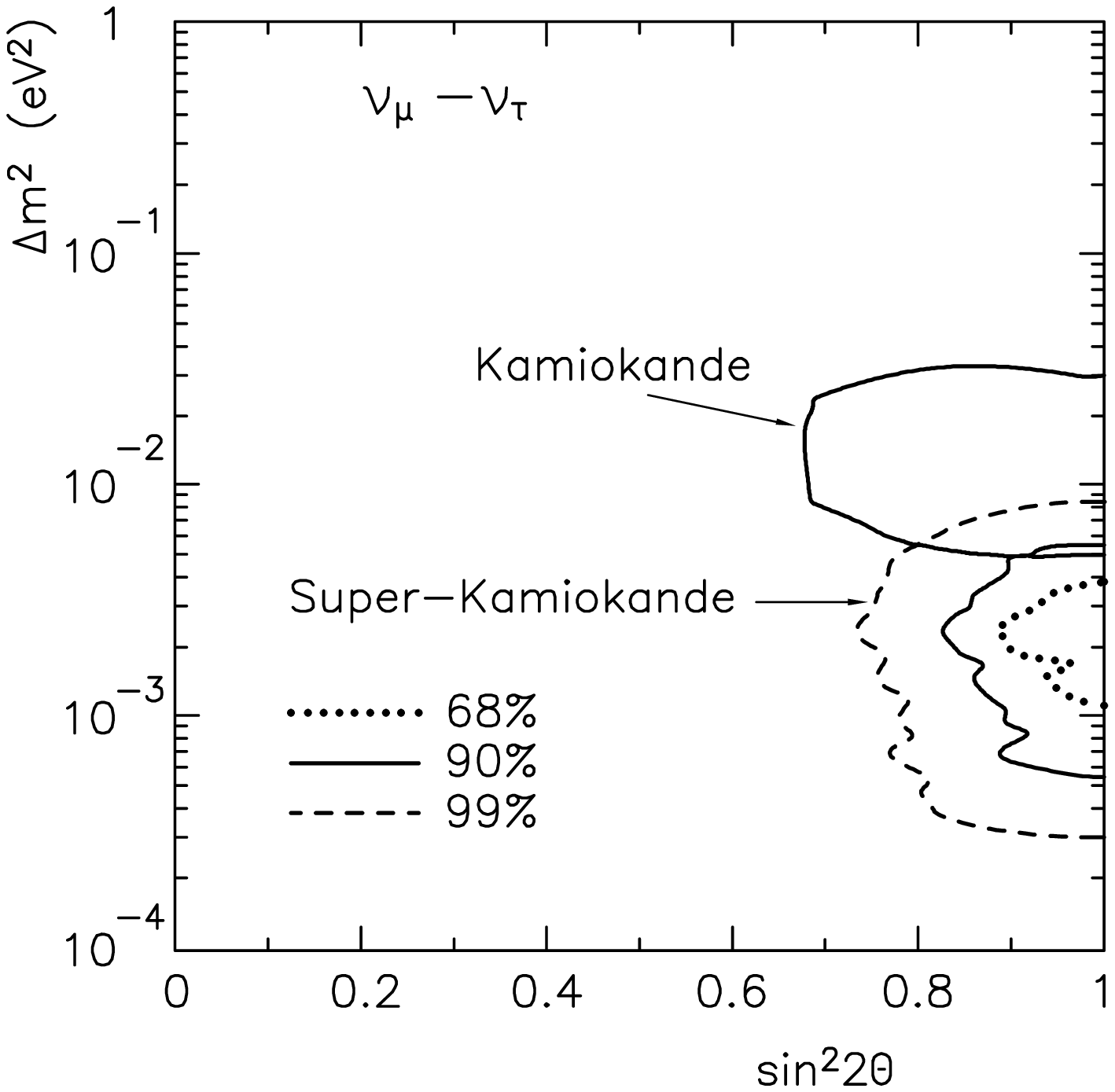,width=6cm}
\end{center}
\caption{Allowed regions in the $(\sin^2 2 \theta ,\delta m^2 )$ plane 
\mbox{for
  $\nu_\mu$ $\leftrightarrow$ $\nu_\tau$ oscillations
from} Ka\-mi\-o\-kan\-de (90\% CL)
  and Super-Kamiokande (68\%, 90\%, 99\% CL) \protect\cite{r7}.}
\label{fig5}
\end{minipage}
\end{figure}

Super-Kamiokande \cite{r6,r7}
 has measured the $\cos\Theta$ dependence not only of $R$,
but also of the $\nu_e$ and $\nu_\mu$ fluxes separately: 
Fig.~\ref{fig4}
shows the $\cos\Theta$ distributions (crosses) of (a) FC $e$-like and (b)
FC $\mu$-like + PC events in the multi-GeV range. (The PC events turned out
to be practically all $\nu_\mu$ events). The rectangles show the MC
predictions for no oscillations. An up-down asymmetry is defined as $A = (U
- D)/(U + D)$ where $U (D)$ is the number of events with
$- 1 < \cos\Theta < -0.2$ ($0.2 < \cos\Theta < 1$). For no oscillations, $A
\approx 0$ is expected for $E_\nu > 1$ GeV, independently of the MC
model. From the data in 
Fig.~\ref{fig4}
the following experimental values were obtained \cite{r6,r7}:
\begin{equation}
\label{e12}
A_e = -0.036 \pm 0.070\, , \ \ \ A_\mu = -0.296 \pm 0.049\, .
\end{equation}
Whereas there is no asymmetry for $\nu_e$, $A_e$ being compatible with
zero, a clear asymmetry with a $6\sigma$ significance is observed for
$\nu_\mu$. Moreover, whereas for $e$-like events the data agree reasonably
well with the MC prediction, for $\mu$-like events a deficit is found at
larger $\Theta$. No explanation other than $\nu$ oscillations could
 be found
for this deficit. An oscillation analysis yielded a much better fit for
$\nu_\mu \leftrightarrow \nu_\tau$ ($\chi^2/NDF = 65/67$) 
than for $\nu_\mu \leftrightarrow \nu_e$ 
($\chi^2/NDF = 88/67$). The best-fit parameters are
\begin{equation}
\label{e13}
\delta m^2 = 2.2 \cdot 10^{-3}\ {\rm eV}^2\, ,\ \ \ 
\sin^2 2 \theta = 1.0\ \ \mbox{for $\nu_\mu \leftrightarrow \nu_\tau$}\, ,
\end{equation}
to be compared
  with (\ref{e11}). 
Fig.~\ref{fig5}
shows in the $(\sin^2 2\theta , \delta m^2)$ plane the allowed regions from
Kamiokande (90\% CL) and from 
Super-Kamiokande (68, 90, 99\% CL). The Kamiokande and 
Super-Kamiokande 90\% CL regions
have only a small overlap around $\delta m^2 \approx 0.5 \cdot 10^{-2}$
eV$^2$. 
With this $\delta m^2$ value, the distance $L_{\rm osc}$ for a full
oscillation (oscillation length $L_{\rm osc} = 4\pi \hbar c E/\delta m^2$)
is $L_{\rm osc} \approx 500$ 
km according to eq.~(\ref{e3}) for $E = 1$ GeV.

In conclusion: Evidence has been found by Kamiokande and 
more significantly by Super-Kamiokande
for $\nu$ oscillations $\nu_\mu \leftrightarrow \nu_\tau$ 
with $\delta m^2 \sim 0.5
\cdot 10^{-2}$ eV$^2$ and maximal mixing. For a hierarchical mass scenario
$m_1 \ll m_2 \ll m_3$ this implies 
\mbox{$m_3 \approx \sqrt{\delta m^2} \sim 0.07$ eV.} 
More generally, this value can be regarded as a lower limit of $m_3$
($m_3 = \sqrt{\delta m^2 + m^2_2} \geq \sqrt{\delta m^2}$).

For completeness we mention two experiments, Kamiokande\cite{b50} and MACRO
\cite{b51}, that have recently measured up-going ($\Theta > 90^\circ$)
muons which pass through the detector from outside. Owing to their large
zenith angle $\Theta$ they cannot be atmospheric muons -- those would not
range so far into the earth --, but are rather produced in CC reactions by
energetic ($\langle E_\nu \rangle \sim 100$ GeV) up-going $\nu_\mu ,
\overline{\nu}_\mu$ in the rock surrounding the detector. In both
experiments a deficit, although not very significant because of large
errors, is observed in the flux of upward through-going muons as 
compared to
the theoretical expectation: $\mu_{\rm obs}/\mu_{\rm MC} = 0.79 \pm 0.18$
from Kamiokande and $0.74 \pm 0.14$ from MACRO; the errors are mainly due
to the relatively large theoretical flux uncertainty. Using the zenith
angular distribution of the upward through-going muons for an oscillation
analysis, Kamiokande obtains the following best-fit parameters in the
physical region (to be compared with the values (\ref{e11}) and
(\ref{e13})): 
\begin{equation}
\label{x1}
\delta m^2 = 3.2 \cdot 10^{-3}\ {\rm eV}^2\, ,\ \sin^2 2\theta = 1.0\ \
{\rm
  for}\ \nu_\mu \leftrightarrow \nu_\tau\, .
\end{equation}
However, the 90\% CL allowed region in the ($\sin^2 2 \theta , \delta m^2$)
plane (not shown) is much larger than the allowed Kamiokande region in
Fig.~\ref{fig5}, which shows that the through-going muon data are much less
restrictive than the data with $\nu$ interactions taking place in the
detector. Similarly, MACRO obtains an allowed region around the values 
\begin{equation}
\label{x2}
\delta m^2 = 2.5 \cdot 10^{-3}\ {\rm eV}^2\, ,\ \sin^2 2\theta = 1.0 
\ \ {\rm for}\ \nu_\mu \leftrightarrow \nu_\tau\, .
\end{equation}

\section{Solar neutrinos}
Solar neutrinos come from the fusion reaction
\begin{equation}
\label{e14}
4p \to {\rm He}^4 + 2e^+ + 2\nu_e
\end{equation}
inside the sun with a total energy release of 26.7 MeV after two $e^+ e^-$
annihilations. The $\nu$ energy spectrum extends up to about 15 MeV with an
average of 
$\langle E_\nu \rangle = 0.59$ MeV. The total $\nu$ flux from the
sun is $\phi_\nu = 1.87 \cdot 10^{38}$ sec$^{-1}$.

\tabcolsep1.5mm
\begin{table}
\caption{The five solar $\nu$ experiments and their results (using a recent
  compilation \protect\cite{r10}). The SSM is BP95.}
\vspace*{0.3cm}
\begin{tabular}{|l|l|l|l|}
\hline
&&Threshold&Result\\
\raisebox{1.5ex}[-1.5ex]{Experiment}&
\raisebox{1.5ex}[-1.5ex]{Reaction}&
[MeV]&
(Result/SSM)\\
\hline\hline
&&&\\[-1ex]
Homestake&Cl$^{37}$$(\nu_e , e^-)$Ar$^{37}$&
$E_\nu > 0.814$& 2.56 $\pm$ 0.22 SNU\\
&&&(0.28 $\pm 0.04$)\\[1ex]
GALLEX&Ga$^{71}$$(\nu_e , e^-)$Ge$^{71}$&
$E_\nu > 0.233$&78 $\pm$ 8 SNU\\
&&&(0.57 $\pm$ 0.05)\\[1ex]
SAGE& Ga$^{71}$$(\nu_e , e^-)$Ge$^{71}$&
$E_\nu > 0.233$&67 $\pm$ 8 SNU\\
&&&(0.49 $\pm$ 0.07)\\[1ex]
Kamiokande&$\nu e \to \nu e$&
$E_\nu > 7.5$&(2.80 $\pm$ 0.38) $\cdot 10^6$ cm$^{-2}$ s$^{-1}$\\
&&&(0.42 $\pm$ 0.09)\\[1ex]
Super-Kam&$\nu e \to \nu e$&
$E_\nu > 6.5$&$\left (2.44{+0.10 \atop -0.09}\right ) 
\cdot 10^6$ cm$^{-2}$ s$^{-1}$\\
&&&(0.37 $\pm$ 0.07)\\
\hline
\end{tabular}
\par\vspace{0.1cm}
{1 SNU (Solar Neutrino Unit)
= 1 $\nu_e$ capture per 10$^{36}$ target nuclei per sec}
\label{t1}
\end{table}

\begin{figure}
\begin{minipage}[b]{5.8cm}
\epsfig{file=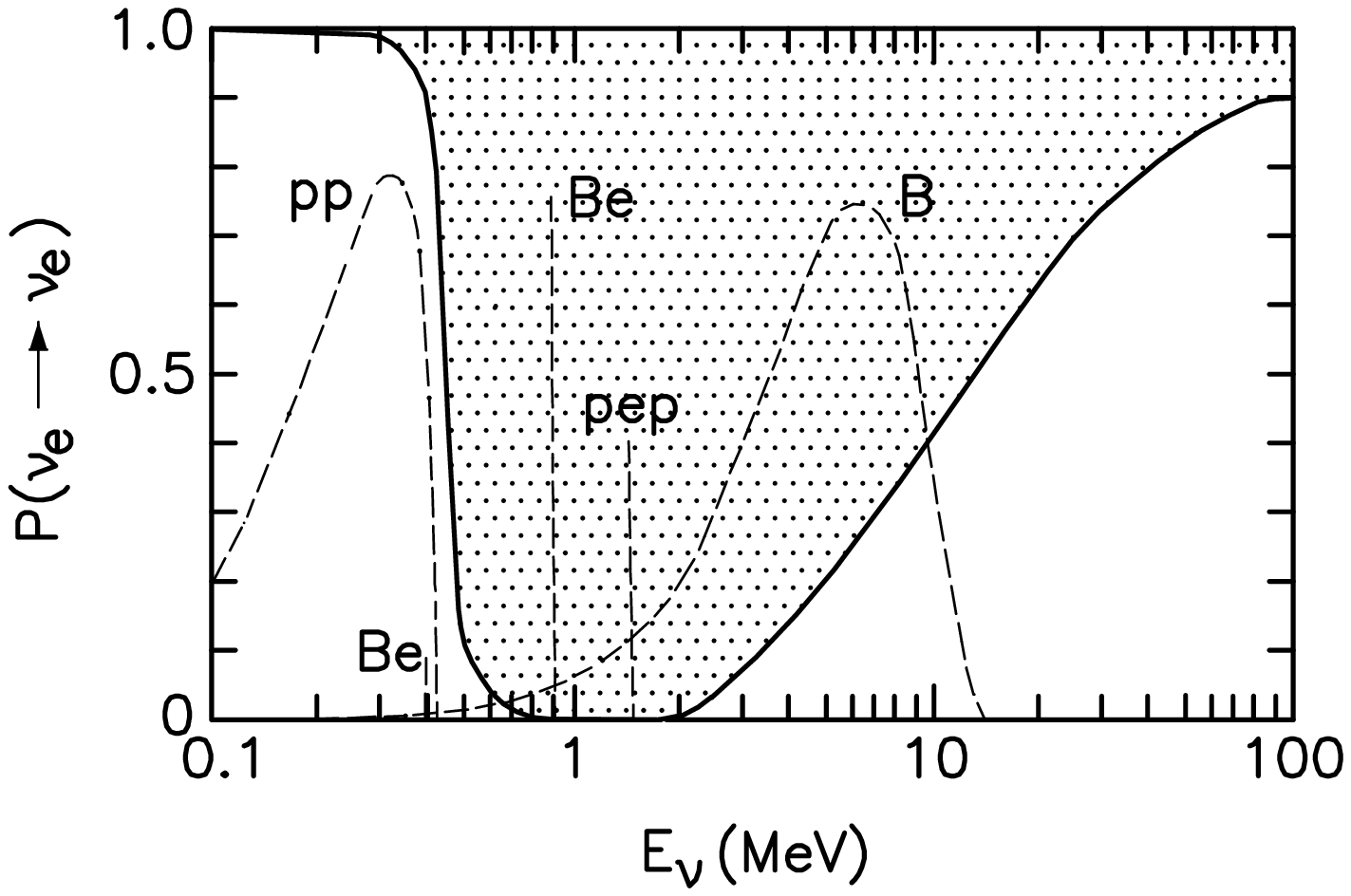,width=5.8cm}
\caption{Survival probability $P (\nu_e \rightarrow \nu_e)$ vs. 
$\nu$ energy $E_\nu$ for
  $\delta m^2 \approx 6 \cdot 10^{-6}$ eV$^2$, $\sin^2 2 \theta \approx
  0.007$ (small-angle solution). The dashed lines show (with arbitrary
  scale) the flux spectra of the pp, Be$^7$, pep and B$^8$ neutrinos.}
\label{figA}
\end{minipage}
\hfill
\begin{minipage}[b]{5.8cm}
\epsfig{file=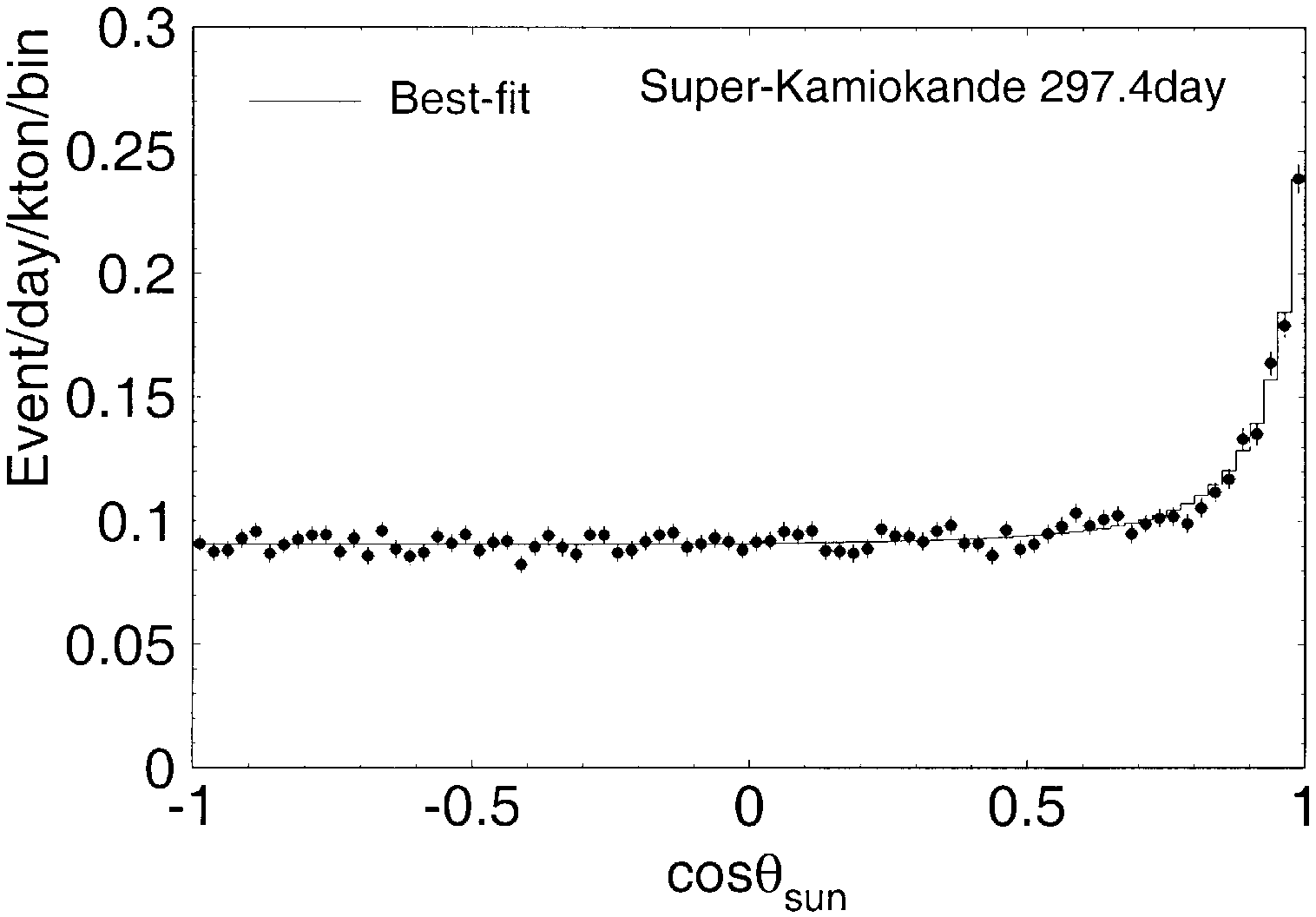,width=5.8cm}
\caption{Distribution of the angle $\theta_{\rm sun}$ between the $e^-$ 
direction and the direction to the sun in events $\nu e \rightarrow \nu e$
from Super-Kamiokande. The line shows the best fit to the data points
\protect\cite{r16}.}
\label{fig6}
\end{minipage}
\end{figure}

Reaction (\ref{e14}) proceeds in various steps in the pp chain or CNO
cycle, the most relevant $\nu_e$ sources being 
\begin{equation}
\label{e15}
\begin{array}{l@{\ :\ \ }l@{\hspace{0.3cm}}l}
pp& p + p \to D + e^+ + \nu_e &(E_\nu < 0.42\ {\rm MeV})\\[1ex]
{\rm Be}^7& {\rm Be}^7 + e^- \to {\rm Li}^7 + \nu_e & 
 (E_\nu = 0.86\ {\rm MeV})\\[1ex]
{\rm B}^8 & {\rm B}^8 \to
{\rm Be}^8 + e^+ + \nu_e & (E_\nu < 14.6\ {\rm MeV})\, .
\end{array}
\end{equation}
$\nu_e$ fluxes from the various sources and rates for the various detection
reactions have been predicted in the framework of the Standard Solar Model
\mbox{(SSM) \cite{r8,r9}.}
 With respect to these predictions a $\nu_e$ deficit from
the sun has been observed by the various experiments as listed in
Tab.~\ref{t1} (see ratios Result/SSM). 
In particular, the Be$^7$-$\nu$ are apparently not registered by the
chlorine and gallium experiments. 
These deficits, the well-known solar
neutrino problem, could be explained by $\nu$ oscillations $\nu_e \to
\nu_X$ into another flavour $X$
($\nu_e$ disappearance) either inside the sun (matter oscillations,
Mikheyev-Smirnov-Wolfenstein (MSW) effect \cite{r11}) with the two possible
solutions \cite{r12}
\begin{equation}
\label{e16}
\begin{array}{l@{\, ,\ }l@{\hspace{0.3cm}}l}
\delta m^2 \approx 5.1 \cdot 10^{-6}\ {\rm eV}^2&
\sin^2 2 \theta \approx 0.0082& \mbox{(small-angle solution)}\\
\delta m^2 \approx 1.6 \cdot 10^{-5}\ {\rm eV}^2&
\sin^2 2 \theta \approx 0.63& \mbox{(large-angle solution)}\, ,
\end{array}
\end{equation}
or on their way from Sun to Earth (vacuum oscillations, $L \approx 1.5
\cdot 10^8$ km) with the solution \cite{r12}
\begin{equation}
\label{e17}
0.5 \cdot 10^{-10} < \delta m^2 < 0.8 \cdot 10^{-10}\ {\rm eV}^2\, ,\ \ 
\sin^2 2 \theta \ \gsim\ 0.65\, .
\end{equation}
As an example, 
Fig.~\ref{figA}
shows the survival probability $P (\nu_e \to \nu_e)$
for the small-angle solution (\ref{e16}) as a function of 
the $\nu$ energy $E_\nu$, together
with the $\nu$ spectra. According to this plot the gallium experiments see
essentially only the pp neutrinos whereas Homestake and 
Kamiokande/Super-Kamiokande see within
their sensitive energy ranges only part of the B$^8$ neutrinos; the Be$^7$
neutrinos  cannot be seen at all by the gallium and chlorine experiments since
they have practically all changed to another flavour, $P (\nu_e \to \nu_e )
\approx 0$. 

After this very brief introduction we report some selected new results
on solar $\nu$ from $\nu$'98. The GALLEX experiment \cite{r13} has been
terminated. The final result from four running periods I to IV with a total
of 65 runs is a $\nu_e$-capture rate of $(78 \pm 8)$ SNU \cite{r14}
(Tab.~\ref{t1}). The successor of GALLEX is GNO (= Gallium Neutrino
Observatory) \cite{r14,r15} in the Gran Sasso Underground Laboratory (Italy).
GNO has
 started data taking in April 1998 with 30 tons of Ga. The experiment
aims at $\sim 60$ tons of Ga by the year 2000 and at $\sim$ 100 tons by
2002. It intends to monitor the solar $\nu$ flux over at least one solar
cycle (11 years) achieving a total error of  less than
5\% ($\sim 4$ SNU) whereas
the present GALLEX errors are 8\% (stat.) and 6\% (syst.).

\begin{figure}
\begin{minipage}[b]{6cm}
\epsfig{file=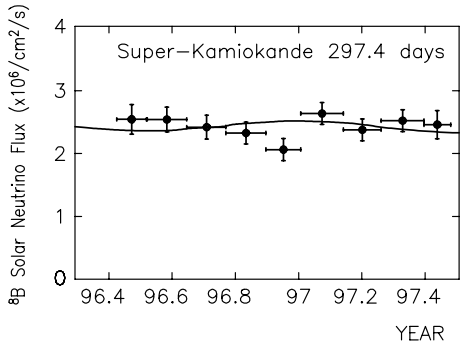,width=6cm}
\end{minipage}
\hfill
\begin{minipage}[b]{5.5cm}
\caption{Time variation of the flux of B$^8$ neutrinos from
  Super-Kamiokande. The line shows the seasonal flux variation expected
  from the excentricity of the earth's orbit around the sun
  \protect\cite{r16}.}
\label{fig7}
\end{minipage}
\end{figure}

Super-Kamiokande \cite{r16} with an energy threshold of 6.5 MeV has
measured the B$^8$-$\nu$ flux 
from the sun via the reaction $\nu + e \to \nu + e$
with a rate of 13.5 solar $\nu$ events per day. For $E_e \gg 1$ MeV
the reaction is strongly
forward peaked, $E_e \theta^2_e < 1$ MeV, where $E_e$
($\theta_e$) is the total energy (scattering angle in rad) of the recoil
electron.
Fig.~\ref{fig6}
shows the $\cos\theta_{\rm sun}$ distribution where $\theta_{\rm sun}$ is
the angle between the direction of the recoil electron and the direction to
the sun. A strong forward peak which is due to solar $\nu$, is observed
above a flat background (coming mainly from $\beta$ decays of
spallation products induced by cosmic-ray muons). 
From the number of events in the peak $(\sim 6800$
at $\nu$'98) a total B$^8$-$\nu$ flux (after correcting for the full SSM
B$^8$-$\nu$ spectrum) of
\begin{equation}
\label{e18}
\phi^{\rm tot} ({\rm B}^8) = \left (2.44 \pm 0.05\ {\rm stat.}\ \pm
{\textstyle{0.09 \atop 0.07}}
\ {\rm syst.}\right ) \cdot 10^6\ {\rm cm}^{-2}\ {\rm
  sec}^{-1}
\end{equation}
is obtained which amounts to $(36.8 \pm 6.5)\%$ of the SSM
(Tab.~\ref{t1}). No seasonal dependence of the solar $\nu$ flux was
observed 
(Fig.~\ref{fig7}),
apart from the expected 6.7\% variation due to the excentricity of the
earth's orbit around the sun ($\varepsilon = 0.0167$, curve in 
Fig.~\ref{fig7},
$\chi^2/NDF = 10/8$). Likewise, 
no significant day-night effect was found,
$(D-N)/(D+N) = -0.023 \pm 0.024$. Such a variation
 could come from a $\nu_e$
regeneration by the MSW effect in the \mbox{earth \cite{r8,r17}.}
Finally, first results were presented on the $E_e$ spectrum for which the
absolute energy scale of Super-Kamiokande
 was calibrated using electrons of known energy
between 5 and 16 MeV from a nearby electron linac \cite{r18}. A precise measurement of
the energy spectrum is of great interest since a distortion of the spectral
shape with respect to the SSM-predicted shape can be caused by $\nu$
oscillations via their energy dependence, see eq.~(\ref{e3}). From a
preliminary analysis vacuum oscillations \cite{r74}
seem to be favoured over matter
oscillations. However, more statistics is needed.

The next forthcoming new solar $\nu$ detectors are SNO (= Sudbury Neutrino
Observatory, Canada) \cite{r19} and Borexino \cite{r20}. SNO (start 1999),
 a heavy-water
Cherenkov detector (1 kton of D$_2$O), will measure the reactions
\begin{equation}
\label{e19}
\begin{array}{l@{\hspace{0.2cm}}l@{\hspace{0.3cm}}l}
\nu_e + D \rightarrow e^- + p + p & \mbox{(CC)} &
E_{\rm thresh} = 1.442\ {\rm MeV}\\
\nu_a + D \rightarrow \nu_a + p + n & \mbox{(NC)} &
E_{\rm thresh} = 2.226\ {\rm MeV}
\end{array}
\end{equation}
($a = e, \mu , \tau$)
above $\sim$ 5 MeV. Whereas the charged-current (CC) reaction is sensitive
only to $\nu_e$, the neutral-current (NC) reaction is sensitive to all
three $\nu$ flavours. Thus by comparing the $\nu$ fluxes obtained from the CC and
NC event rates one will find out directly whether or not the solar $\nu$ deficit is
due to $\nu$ oscillations $\nu_e \to \nu_X$, while the total solar $\nu$
flux (independent of flavour) agrees with the SSM prediction. 
Borexino (start 2000), an
organic liquid scintillator detector (300 tons) in the Gran Sasso
laboratory, will measure the reaction $\nu + e \to \nu + e$ with a very low
threshold of $\sim$ 0.25 MeV, thus covering a large part of the $E_e$
spectrum from the
so far unseen
Be$^7$-$\nu$. Since the reaction is sensitive to all three $\nu$ flavours
(although with different cross sections for $\nu_e e$ and for $\nu_\mu e$,
$\nu_\tau e$ scattering) Borexino will allow to decide whether the original
Be$^7$-$\nu_e$ have only changed their flavour or whether 
Be$^7$-$\nu$ are missing from the very beginning which would be in conflict
with the SSM.

\section{LSND and KARMEN}
\begin{figure}
\begin{minipage}[b]{7cm}
\epsfig{file=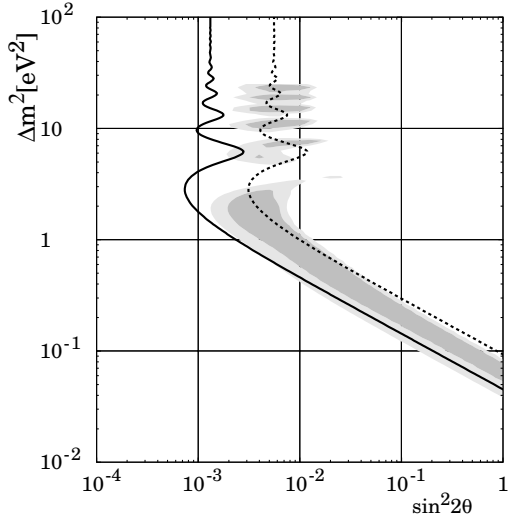,width=7cm}
\end{minipage}
\hfill
\begin{minipage}[b]{4.8cm}
\caption{The shaded areas show the 90\% CL (darkly shaded) and 99\% CL
  (lightly shaded) allowed parameter region from LSND for $\nu_\mu 
\leftrightarrow
  \nu_e$ oscillations. The region to the
  right of the full curve is excluded at 90\% CL by KARMEN 2. The dotted
  line is the present sensitivity of KARMEN 2. From J.~Kleinfeller
  \protect\cite{r40}.}
\label{figB}
\end{minipage}
\end{figure}
The LSND experiment at Los Alamos (LSND = Liquid 
Scintillator Neutrino
Detector) has previously observed \cite{r21} events of the type 
$\overline{\nu}_e p \rightarrow n e^+$ in connection with $\pi^+$ 
decays at rest
and subsequent $\mu^+$ decays at rest ($\pi^+ \to \mu^+ \nu_\mu$,
$\mu^+ \to e^+ \nu_e \overline{\nu}_\mu$) from which no 
$\overline{\nu}_e$ are expected. The occurrence of $17.4 \pm 4.7$ 
$\overline{\nu}_e p$ events ($\overline{\nu}_e$ appearance) above
background with $36 < E_e < 60$ MeV was attributed to flavour transitions
$\overline{\nu}_\mu \to \overline{\nu}_e$ of $\overline{\nu}_\mu$
from $\mu^+$ decays. The allowed regions in the $(\sin^2 2\theta , \delta
m^2)$ plane from an oscillation analysis are shown by the shaded areas in 
Fig.~\ref{figB}.
In the meantime the signal above background has gone up from 17.4 to 20.8
$\pm$ 5.4 excess events \cite{r22} (1993-1997 data), yielding an
oscillation probability of $(0.31 \pm 0.09 \pm 0.05$)\% (preliminary). The
evidence for $\nu$ oscillations is strengthened by observing 
\cite{r22,r23} in
the same experiment also an excess of 18 $\nu_e$ events 
($\nu_e + C \to
e^- +X$) above the expected $\nu_e$ 
background from $\mu^+ \to e^+ \nu_e
\overline{\nu}_\mu$ and $\pi^+ \to e^+ \nu_e$. These excess events are
attributed to $\nu_\mu \to \nu_e$ transitions, the $\nu_\mu$ coming from
$\pi^+ \to \mu^+ \nu_\mu$ decays in flight, and lead to the same allowed
regions in the $(\sin^2 2\theta , \delta
m^2)$ plane as the $\overline{\nu}_e p$ events.

The KARMEN Collaboration (= KArlsruhe-Rutherford Medium Energy Neutrino
experiment) \cite{r24} is carrying out with an upgraded detector (KARMEN 2)
an experiment with stopped $\pi^+$ and $\mu^+$,
similar to LSND and using the same reactions for detecting
$\overline{\nu}_e$ or $\nu_e$. So far no $\overline{\nu}_e p$ event has
been found where $2.88 \pm 0.13$ background events are expected. The 
$(\sin^2 2\theta , \delta
m^2)$ region which is excluded by KARMEN at 90\% CL (full curve in
Fig.~\ref{figB}), 
covers almost the total allowed
region of LSND. So the two experiments are hardly compatible. KARMEN is
taking more data to increase their sensitivity (dotted curve in 
Fig.~\ref{figB}).

An experiment (BOONE = BOOster Neutrino Experiment) \cite{r22} is
forthcoming at the booster accelerator of Fermilab to definitely verify or
refute the LSND result and to measure the oscillation parameters. The
experiment has been approved and is foreseen to start by 2001.

\section{Possible neutrino mass schemes}
Table \ref{tab2} summarizes the results on $\delta m^2_{ij}$ obtained from
various oscillation analyses, considering only those experiments that have
apparently observed a positive oscillation signal.
 All reactor and accelerator
experiments, except LSND, have so far found no indications for $\nu$
oscillations thus yielding only limits on $\delta m^2$ and $\sin^2
2\theta$. The table includes the prediction for the sum of the light $\nu$
masses which comes from a cosmological model with non-baryonic cold and hot
dark matter (CHDM), both needed to explain the 
observed structure and density in the
universe at all distance scales \cite{rx3}: 
Assuming a flat universe without
cosmological constant ($\Omega = 1$, $\Lambda = 0$) one obtains with 20\%
hot dark matter in the form of light relic neutrinos ($\sum_\nu \Omega_\nu
= 0.2$):
\begin{equation}
\label{e20}
\sum_\nu m_\nu = 94 \cdot \sum_\nu \Omega_\nu h^2_0\ {\rm eV}\ \approx 4.7\
{\rm eV}\, ,
\end{equation}
where $h_0 \approx 0.5$ is the normalized Hubble constant ($H_0 = 100\ h_0$
km s$^{-1}$ Mpc$^{-1}$).

\begin{table}
\caption{Allowed ranges for $\delta m^2_{ij}$ from the various oscillation
  analyses and constraint on the $\nu$ masses from a cold-hot dark matter
  model}
\vspace*{0.2cm}
\begin{center}
\begin{tabular}{|l|l|l|}
\hline
&&\\[-1ex]
Measurement&$|\delta m^2_{ij}|$ (eV$^2$)&Oscillation\\[0.5ex]
\hline\hline
&&\\[-1ex]
atmospheric $\nu$&$(0.4 - 7) \cdot 10^{-3}$&
$\nu_\mu \leftrightarrow \nu_\tau$\\[1ex]
solar $\nu$& &$\nu_e \leftrightarrow \nu_\mu , \nu_s$\\
\ \ -- matter&$(0.5 - 1.6) \cdot 10^{-5}$&\\
\ \ -- vacuum&$(0.5 - 0.8) \cdot 10^{-10}$&\\[1ex]
LSND&$0.2 - 10$&$\nu_\mu \leftrightarrow \nu_e$\\
\hline
hot dark matter&$\sum_\nu m_\nu \approx 5$ eV&\\
\hline
\end{tabular}
\end{center}
\label{tab2}
\end{table}

\begin{figure}
\begin{minipage}[b]{6cm}
\epsfig{file=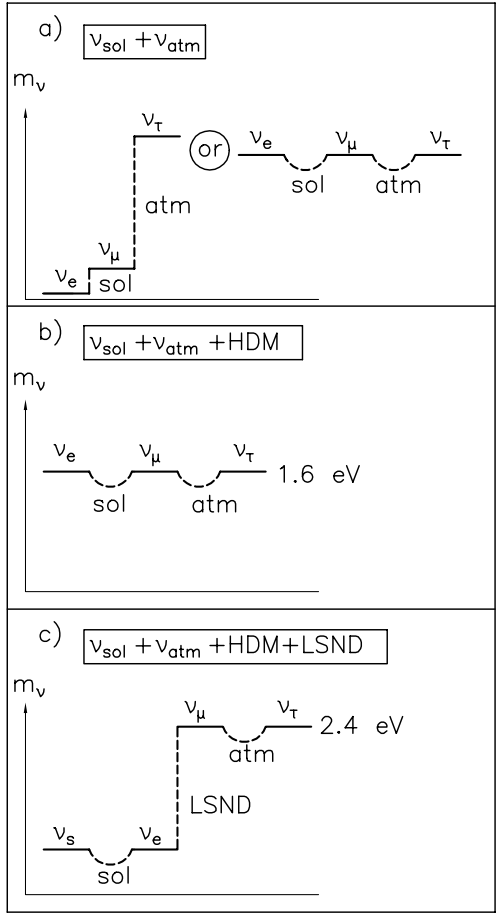,width=6.5cm}
\end{minipage}
\hfill
\begin{minipage}[b]{5cm}
\caption{Sketches showing qualitatively some possible $\nu$ mass
  assignments in order to accommodate (a) the solar $\nu$ and atmospheric
  $\nu$ results, (b) the solar $\nu$, atmospheric $\nu$ results and hot
  dark matter (HDM), and (c) the solar $\nu$, atmospheric $\nu$, LSND
  results and HDM.}
\label{fig8}
\end{minipage}
\end{figure}

From Tab.~\ref{tab2} it seems evident that there are three independent 
$\delta m^2_{ij}$ values. This requires a minimum of four different $\nu$
species since for three $\nu$ flavours the three possible 
$\delta m^2_{ij}$ are not independent: $\delta m^2_{12} + \delta m^2_{23}
+ \delta m^2_{31} = 0$. The fourth $\nu$ must be a sterile (inactive)
neutrino ($\nu_s$), which does not participate in the weak interaction
(e.g.\ a right-handed $\nu$), since from the LEP experiments the number of
light active neutrinos, coupling to $Z^0$, is known to be 3
\mbox{(2.993 $\pm$ 0.011) \cite{r25}.} An alternative,
leaving it at three $\nu$ species, is of course that one of the 
measurements
in Tab.~\ref{tab2} is wrong. We now discuss, step by step, some
possible mass assignments in a qualitative way (Fig.~\ref{fig8}) 
\cite{r26}; we simplify the presentation somewhat by assuming -- against large-mixing
angle solutions --  that the mixing angles $\theta_{e\mu}$,
$\theta_{\mu\tau}$, $\theta_{\tau e}$ are small. In this case $\nu_1 ,
\nu_2 , \nu_3$ are the dominant states in $\nu_e$, $\nu_\mu$, $\nu_\tau$,
respectively, such that $m (\nu_e) \approx m_1$, $m (\nu_\mu) \approx m_2$,
$m (\nu_\tau) \approx m_3$ according to eq.~(\ref{e4}), i.e.\ the measured
$\delta m^2_{ij}$ for the mass eigenstates hold approximately also for the flavour
eigenstates. Quantitative treatments can be found 
elsewhere \cite{r26,r27}.

The atmospheric and solar results alone can be accommodated in two ways
(Fig.~\ref{fig8}a): 
\begin{itemize}
\item
Three $\nu$ with a hierarchical mass pattern, $m (\nu_e) \ll m (\nu_\mu)
\ll m (\nu_\tau)$, as predicted by the seesaw mechanism. Using the $\delta
m^2$ values in Tab.~\ref{tab2}, this pattern implies $m (\nu_e) \approx 0$,
$m (\nu_\mu) \approx 3 \cdot 10^{-3}$ or $10^{-5}$ eV (from solar $\nu$)
and $m (\nu_\tau) \approx 7 \cdot 10^{-2}$ eV (from atmospheric $\nu$). In
this scenario there is no room for e.g.\ the $\nu_\tau$ being the hot dark matter
particle.
\item
Three $\nu$ with a democratic (nearly degenerate) mass pattern, $m (\nu_e)
\approx m (\nu_\mu) \approx m (\nu_\tau)$, with 
relatively small mass differences such
that they yield the experimental $\delta m^2$ values from solar and
atmospheric $\nu$ oscillations respectively.
\end{itemize}

This latter mass pattern is still satisfactory if we add the CHDM
prediction (\ref{e20}) 
as a further constraint. In this case (Fig.~\ref{fig8}b) $m
(\nu_e) \approx m (\nu_\mu) \approx m (\nu_\tau) \approx 1.6$ eV such that
$\sum_\nu m_\nu \approx 4.8$ eV.

Finally, if also the LSND result is added, a fourth, sterile $\nu$ 
$(\nu_s)$
is needed, as explained above. A possible mass pattern is shown in 
Fig.~\ref{fig8}c: Two rather light, nearly degenerate neutrinos, $\nu_e$
and $\nu_s$, and two somewhat heavier, also nearly degenerate neutrinos,
$\nu_\mu$ and $\nu_\tau$, as hot dark matter  particles with $m (\nu_\mu) \approx m
(\nu_\tau) \approx 2.4$ eV so that constraint (\ref{e20}) 
is fulfilled. In this
scheme the solar $\nu_e$ deficit comes from $\nu_e \leftrightarrow \nu_s$
oscillations, the atmospheric $\nu_\mu$ deficit from $\nu_\mu
\leftrightarrow \nu_\tau$ oscillations and the LSND result from $\nu_e
\leftrightarrow \nu_\mu$ oscillations with a relatively large $\delta m^2
\approx (2.4$ eV)$^2 = 5.8$ eV$^2$. An alternative mass arrangement is to
interchange $\nu_s$ and $\nu_\tau$ in Fig.~\ref{fig8}c.

It has been claimed that the present atmospheric, solar and LSND results
can also be described \cite{r70} by three $\nu$ flavours, i.e.\ by two
independent $\delta m^2$ values, if one applies the three-flavour
oscillation formalism to all three 
experimental results simultaneously, instead of
employing the two-flavour formalism, eq.~(\ref{e3}), for each result
separately. See however ref.~\cite{r71} where the consequences of the CHOOZ
result (see below) are analyzed.

\section{Some further results and future plans}
Two short-baseline (SBL) accelerator experiments, CHORUS \cite{r28} and
NOMAD \cite{r29}, have been searching for 
$\nu_\mu \to \nu_\tau$ transitions
in a wide-band $\nu_\mu$ beam at the CERN SPS ($\langle E_\nu \rangle
\approx 27$ GeV, $L \sim 800$ m, i.e.\ $\langle E_\nu \rangle /L \sim 7$
eV$^2$) by looking for events $\nu_\tau N \to \tau^- X$. No $\nu_\tau$
event was found so far in either experiment from which the following 90\% CL upper
limits on the mixing angle at large $\delta m^2$ were deduced: $\sin^2 2
\theta_{\mu\tau} < 1.8 \cdot 10^{-3}$ (CHORUS), $< 4.2 \cdot 10^{-3}$
(NOMAD).

\begin{figure}
\begin{minipage}[b]{6cm}
\epsfig{file=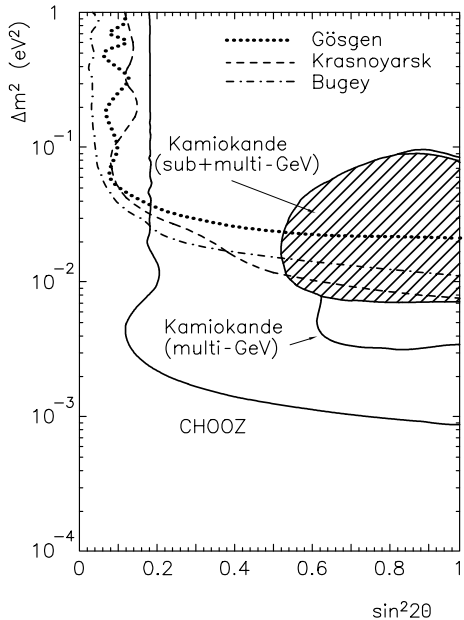,width=6cm}
\end{minipage}
\hfill
\begin{minipage}[b]{5.5cm}
\caption{Exclusion plot of CHOOZ for $\overline{\nu}_e \to
  \overline{\nu}_X$; the region to the right of the CHOOZ curve is excluded
  at 90\% CL. Shown are also the 90\% CL exclusion curves from three SBL
  reactor experiments and the 90\% CL allowed region of Kamiokande for
  $\nu_\mu \to \nu_e$ from atmospheric $\nu$ \protect\cite{r30}.}
\label{fig9}
\end{minipage}
\end{figure}

Long-baseline (LBL) experiments (large $L/E$) are according to 
(\ref{e3}) sensitive to smaller $\delta m^2$. The first LBL reactor
experiment has been carried out by the CHOOZ collaboration \cite{r30} at a
nuclear power station near Chooz in France ($\langle E_\nu \rangle \sim 3$
MeV, $L = 1$ km, i.e.\ $\langle E_\nu \rangle /L \sim 6 \cdot 10^{-4}$
eV$^2$), searching for $\overline{\nu}_e \to \overline{\nu}_X$
disappearance. No evidence for $\nu$ oscillations was found, the 
two-flavour analysis
yielding the exclusion plot shown in Fig.~\ref{fig9}. The $(\sin^2 2
\theta , \delta m^2)$ region to the right of the CHOOZ curve is excluded at
90\% CL. In particular the following upper limits were obtained from the plot:
$\delta m^2 < 0.9 \cdot 10^{-3}$ eV$^2$ for $\sin^2 2 \theta = 1$ (maximum
mixing), $\sin^2 2 \theta < 0.18$ for large $\delta m^2$. The $\delta m^2$
limit improves the limits obtained from previous SBL reactor experiments,
included in Fig.~\ref{fig9}, by about an order of magnitude. The figure
also shows that the CHOOZ result excludes the allowed region which
Kamiokande has obtained for $\nu_\mu \leftrightarrow \nu_e$ oscillations from their
atmospheric $\nu$ measurements. The corresponding allowed region of
Super-Kamiokande is excluded by CHOOZ to a large part.

With accelerators the following three LBL experiments are being prepared
for the near future \cite{r31}:
\begin{itemize}
\item
In Japan a beam from KEK to Super-Kamiokande (K2K \cite{r32}) with $\langle
  E_\nu \rangle \sim 1$ GeV and $L = 250$ km, i.e.\ $\langle E_\nu \rangle
  /L \sim 10^{-3}$ eV$^2$. The experiment will look for $\nu_\mu$
  disappearance and $\nu_e$ appearance. K2K is expected to start data
  taking in 1999.
\item
In USA a $\nu_\mu$ beam from Fermilab to Soudan 2 
(E875/MINOS \cite{r33}) with
$\langle E_\nu \rangle \approx 11$ GeV and $L = 730$ km, i.e.\  
$\langle E_\nu \rangle /L \sim 3 \cdot 10^{-3}$ eV$^2$. MINOS is expected
to start data taking in 2002.
\item
In Europe a $\nu_\mu$ beam \cite{r41}
from CERN to the Gran Sasso laboratory with
$\langle E_\nu \rangle \sim 25$ GeV and $L = 730$ km, i.e.\ 
$\langle E_\nu \rangle /L \sim 7 \cdot 10^{-3}$ eV$^2$.
Three detectors are planned:
Two detectors, 
ICARUS \cite{r34}, a liquid-argon drift chamber, and OPERA \cite{r39}, 
employing a
target with emulsion sheets (similar to CHORUS), will search
 for $\nu_\tau$ appearance; the third detector, NOE \cite{r72}, a
fine-grained massive calorimeter based on scintillating fiber technique,
will search for $\nu_\mu$ disappearance as well as for $\nu_e$, $\nu_\tau$
appearance.  
\end{itemize}
In all these experiments,
given the baseline $L$, $E_\nu$ should be on the one hand as small as
possible in order to reach low $\delta m^2$. On the other hand the event
rate in the detector decreases with decreasing $E_\nu$ for two reasons: (a)
the cross section is proportional to $E_\nu$, and (b) the divergence of the
$\nu$ beam gets larger with smaller $E_\nu$. The beam divergence is a
severe limitation at large distances $L$ and calls for a correspondingly
large detector mass and/or high $\nu$ beam intensity. The latter could be
achieved by using neutrinos not, as usually, from 
charged $\pi$ and $K$ decays, but
from $\mu$ decays in future high-luminosity muon storage rings with long
straight sections pointing in the desired direction \cite{r73}.

Two new direct measurements of upper limits to the $\nu_e$ mass from
tritium $\beta$-decay, H$^3 \to$ He$^3 + e^- + \overline{\nu}_e$, were
presented at $\nu$'98:
\begin{equation}
\label{e21}
\begin{array}{l@{\, :\ \ }l}
\mbox{Mainz experiment \cite{r35}}&
m^2_\nu = (-9 \pm 8 \pm 2)\ {\rm eV}^2 \Rightarrow m_\nu < 3.4\ {\rm eV}
\\[1ex]
\mbox{Troitsk experiment \cite{r36}}&
m^2_\nu = (-2.1 \pm 3.7 \pm 2.3)\ {\rm eV}^2 
\Rightarrow m_\nu < 2.7\ {\rm eV}\, ,
\end{array}
\end{equation}
where the upper $m_\nu$ limits are at 95\% CL.

Finally, the Heidelberg-Moscow collaboration \cite{r37}, searching in the
Gran Sasso laboratory with a germanium detector for the neutrinoless double
$\beta$-decay ($0\nu\beta\beta$ decay) of Ge$^{76}$, Ge$^{76} \to$ Se$^{76}
+ 2e^-$, has determined the following new 90\% CL limits for the half-life
and the (average) $\nu$ mass (provided that the $\nu_e$ is a Majorana
neutrino, i.e.\ $\nu^M \equiv \overline{\nu^M}$!):
\begin{equation}
\label{e22}
T^{0\nu}_{\frac{1}{2}} ({\rm Ge}^{76}) > 1.1 \cdot 10^{25}\
 {\rm years}\ \Rightarrow
\langle m^M_\nu \rangle < 0.46\ {\rm eV}\, .
\end{equation}
For the longer range future a detector consisting eventually of $\sim 1$
ton of Ge (GENIUS \cite{r38}) is envisaged which is to be realized in
several steps.

\section*{Acknowledgments}
I am grateful to the organizers of the Delphi symposium for a
fruitful and enjoyable meeting with an excellent scientific and social
program and many lively discussions. I wish to thank Mrs.~Edeltraud Haag
for typing the text in \LaTeX\ and arranging the figures.

\end{document}